\documentstyle[amssymb,aps,epsf]{revtex}
\begin{document}
\title{Can Reissner-Nordstr\"{o}m wormholes be considered for Spacetime Foam
formation?}
\author{Remo Garattini}
\address{Facolt\`{a} di Ingegneria, Universit\`{a} degli Studi di Bergamo,\\
Viale Marconi, 5, 24044 Dalmine (Bergamo) Italy.\\ E-mail:
Garattini@mi.infn.it}
\maketitle
\begin{abstract}
A simple model of spacetime foam, made by $N$
Reissner-Nordstr\"{o}m wormholes with a magnetic and electric
charge in a semiclassical approximation, is taken under
examination. The Casimir-like energy of the quantum fluctuation of
such a model is computed and compared with that one obtained with
a foamy space modeled by $N$ Schwarzschild wormholes. The
comparison leads to the conclusion that a foamy spacetime cannot
be considered as a collection of $N$ Reissner-Nordstr\"{o}m
wormholes but that such a collection can be taken as an excited
state of the foam.
\end{abstract}
\section{Introduction}
\label{p0}Among the long-standing problems in physics, a fundamental one,
from a theoretical point of view, is the absence of a quantum theory of
gravity. Even if a lot of work has been devoted, especially in the direction
of string theory and subsequent modifications, i.e. the branes, a complete
scheme is still lacking. On the other hand in the traditional path integral
approach to quantum gravity, a typical problem is that the generating
functional
\begin{equation}
\int {\cal D}\left[ g_{\mu \nu }\right] \exp iS_{g}\left[ g_{\mu \nu }\right]
\label{p01}
\end{equation}
is ill defined because ${\cal D}\left[ g_{\mu \nu }\right] $ does not
represent a measure and moreover there is no convergence. However, in the
context of a WKB approximation with a fixed background, one can obtain
interesting informations. Indeed, if one considers a background $\bar{g}%
_{\mu \nu }$, the gravitational field splits into
\begin{equation}
g_{\mu \nu }=\bar{g}_{\mu \nu }+h_{\mu \nu },
\end{equation}
where $h_{\mu \nu }$ is a quantum fluctuation around the
background field. Then Eq. $\left( \ref{p01}\right) $ becomes
\begin{equation}
\int {\cal D}g_{\mu \nu }\exp iS_{g}\left[ g_{\mu \nu }\right] \simeq \exp
iS_{g}\left[ \bar{g}_{\mu \nu }\right] \int {\cal D}h_{\mu \nu }\exp
iS_{g}^{\left( 2\right) }\left[ h_{\mu \nu }\right] ,
\end{equation}
where $S_{g}^{\left( 2\right) }\left[ h_{\mu \nu }\right] $ is the action
approximated to second order. In the context of the Euclidean path
integration, last expression becomes
\begin{equation}
\int {\cal D}g_{\mu \nu }\exp \left( -I_{g}\left[ g_{\mu \nu }\right]
\right) \simeq \exp \left( -I_{g}\left[ \bar{g}_{\mu \nu }\right] \right)
\int {\cal D}h_{\mu \nu }\exp \left( -I_{g}^{\left( 2\right) }\left[ h_{\mu
\nu }\right] \right) .
\end{equation}
Previous equation can be cast in the form
\begin{equation}
\Gamma =A\exp \left( -I_{cl}\right) ,  \label{p02}
\end{equation}
where $\Gamma $ is the decay probability per unit volume and time,
while $A$ is the prefactor coming from the saddle point evaluation
and $I_{cl}$ is the classical part of the action. If a single
negative eigenvalue appears in the prefactor $A$, it means that
the related bounce shifts the energy of the false ground
state\cite{Coleman}. In particular, it is possible to discuss
decay probabilities from one spacetime to another one \cite
{BoHaw,GPY,GP,Young,VW,Prestidge}. For certain classes of
gravitational backgrounds, namely the static spherically symmetric
metrics, it could be interesting the use of other methods based on
variational approach. In a series of papers, we have used such an
approach to show that a model of spacetime foam can be concretely
realized if one considers a collection of Schwarzschild wormholes
whose energy is given by a Casimir energy\cite
{Remo,Remo1,Remo2,Remo3,Remo4}. We recall that the Casimir energy
involves a subtraction procedure between zero point energies
having the same boundary conditions. In this paper we would like
to apply such methods to another class of static spherically
symmetric wormholes involving the electromagnetic field. This
class is described by the Reissner-Nordstr\"{o}m (RN) metric which
is a solution of the Einstein-Maxwell equations with two basic
parameters: a mass $M$ and a charge $Q$. Together with the
Schwarzschild metric also the RN metric shares the property of
being asymptotically flat. This means that the Casimir energy
computation for such wormholes class will be made with flat space
as a reference space. The final expression will be compared with
that one obtained for the Schwarzschild wormholes showing that the
Casimir energy for RN wormholes is always higher than the Casimir
energy for the Schwarzschild wormholes. This means that RN
wormholes cannot be taken as a representation of the ground state
of a foamy spacetime. To this purpose we will fix our attention to
the following quantity
\begin{equation}
E^{RN}\left( M,Q\right) =E^{flat}\left( 0\right) +\Delta E_{flat}^{RN}\left(
M,Q\right) _{|classical}+\Delta E_{flat}^{RN}\left( M,Q\right) _{|1-loop},
\label{i0}
\end{equation}
representing the total energy computed to one-loop in a RN background. $%
E^{flat}\left( 0\right) $ is the reference space energy, which in
this case is flat space. $\Delta E_{flat}^{RN}\left( M,Q\right)
_{|classical}$ is the energy difference between the RN and the
flat metrics, stored in the boundaries and $\Delta
E_{flat}^{RN}\left( M,Q\right) _{|1-loop}$ is the quantum
correction to the classical term. The analogy between $\Delta
E_{flat}^{RN}\left( M,Q\right) _{|1-loop}$ and the prefactor of
Eq.$\left( \ref{p02}\right)$ is the key to obtain information
about instability. Indeed if we discover that the spectrum of the
second order differential operator associated with the quantum
fluctuations of transverse and traceless tensors (TT) admits bound
states, then we have instability. In this paper we will assume
that only one unstable mode appears in the spectrum of TT tensors;
this is sufficient to apply Coleman arguments about transition
from a vacuum to another one. Note that with this assumption we
are in the worst situation, namely the Schwarzschild and RN
wormholes can be compared. In fact without the negative mode a
spontaneous transition from vacuum to vacuum cannot happen.
Therefore in this framework, it is sufficient to compute the
stable part of the Casimir energy to compare these different
pictures. To concretely compute Eq.$\left( \ref{i0}\right) $ we
refer to the following Hamiltonian with boundary
\begin{equation}
H_{T}=H_{\Sigma }+H_{\partial \Sigma }=\int_{\Sigma }d^{3}x\left( N{\cal H+}%
N_{i}{\cal H}^{i}\right) +H_{\partial \Sigma },
\end{equation}
where $N$ is the {\it lapse} function, $N_{i}$ is the {\it shift }function
and
\begin{equation}
\left\{
\begin{array}{l}
{\cal H}{\bf =}G_{ijkl}\pi ^{ij}\pi ^{kl}\left( \frac{16\pi G}{\sqrt{g}}%
\right) -\frac{\sqrt{g}}{16\pi G}R^{\left( 3\right) }+{\cal H}_{M} \\
{\cal H}^{i}=-2\pi _{|j}^{ij}+{\cal H}_{M}^{i}.
\end{array}
\right.
\end{equation}
${\cal H}_{M}$ is the energy-momentum tensor contribution of the
electromagnetic field. $H_{\partial \Sigma }$ represents the energy stored
into the boundary. Since the metric considered has no off-diagonal elements,
the Hamiltonian becomes
\begin{equation}
H_{T}=H_{\Sigma }+H_{\partial \Sigma }=\int_{\Sigma }d^{3}xN{\cal H}%
+H_{\partial \Sigma }.
\end{equation}
However the physical quantity of interest is the Casimir energy.
Therefore we will consider the following expectation value
\begin{equation}
\Delta E_{flat}^{RN}\left( M,Q\right) =\frac{\left\langle \Psi \left|
H_{\Sigma }^{RN}-H_{\Sigma }^{flat}\right| \Psi \right\rangle }{\left\langle
\Psi |\Psi \right\rangle },
\end{equation}
where $H_{\Sigma }^{RN}$ and $H_{\Sigma }^{flat}$ are the total Hamiltonians
referred to the RN and flat spacetimes respectively for the volume term\cite
{Remo,Remo5,Remo6} and $\Psi $ is a {\it trial wave functional} of the
gaussian form. Note that the flat space is a particular case of the RN space
with the parameters $M=Q=0$. Since the reference space is the same one of
the Schwarzschild space, it is immediate to recognize that if an instability
appears, flat space could decay via a Schwarzschild black hole pair creation
or via a Reissner-Nordstr\"{o}m black hole pair creation. Therefore to give
indications about the ``{\it ground state''} of a foamy spacetime, it is
important to establish if
\begin{equation}
\Delta E_{flat}^{RN}\left( M,Q\right) \lessgtr \Delta
E_{flat}^{Schwarzschild}\left( M\right)
\end{equation}
to one-loop approximation. It is important to remark that it is the vacuum
energy difference with the same asymptotic reference space that gives the
possibility to choose which vacuum is appropriate and not the direct energy
difference between these possible candidates. The rest of the paper is
structured as follows: in section \ref{p1}, we introduce the
Reissner-Nordstr\"{o}m metric; in section \ref{p2}, we compute the boundary
energy by means of quasilocal energy; in section \ref{p3}, a variational
calculation will be set up to compute the Casimir energy and in particular
we will restrict our analysis to the transverse-traceless tensor metric
fluctuations (TT); in section \ref{p4}, we evaluate the spectrum of the
operator associated with TT tensors (Laplace-Beltrami operator) and we
compare the result with what we have obtained in case of Schwarzschild
wormholes; in section \ref{p5}, we summarize and conclude. Units in which $%
\hbar =c=k=1$ are used throughout the paper.
\section{The Reissner-Nordstr\"{o}m metric}
\label{p1}The RN line element is
\begin{equation}
ds^{2}=-f\left( r\right) dt^{2}+f\left( r\right) ^{-1}dr^{2}+r^{2}d\Omega
^{2},  \label{p11}
\end{equation}
with
\begin{equation}
f\left( r\right) =\left( 1-\frac{2MG}{r}+\frac{Q^{2}}{r^{2}}\right) ,
\label{p12}
\end{equation}
where $Q^{2}=G\left( Q_{e}^{2}+Q_{m}^{2}\right) $; $Q_{e}$ and $Q_{m}$ are
the electric and magnetic charge respectively. When the electric charge is
considered the electromagnetic potential assumes the form $A_{\alpha
}=\left( Q_{e}/r,0,0,0\right) $ and the electromagnetic tensor $F_{\alpha
\beta }=\partial _{\alpha }A_{\beta }-\partial _{\beta }A_{\alpha }$ is $%
F_{01}=-Q_{e}/r^{2}$. In the case of a pure magnetic field, the form is $%
A_{\alpha }=\left( 0,-Q_{m}\sin \theta ,0,0\right) $ and the electromagnetic
tensor becomes $F_{23}=$ $Q_{m}\sin \theta $. Therefore, although the
gravitational potential $f\left( r\right) $ assumes the same form, the
gravitational perturbation contribute in a different way. When $Q=0$ the
metric describes the Schwarzschild metric. When $Q=M=0$, the metric is flat.
For $Q\neq 0$, we can distinguish three different cases:
\begin{itemize}
\item[a)]  $MG>Q$. In this case the gravitational potential $f\left(
r\right) $ admits two real distinct solutions located at
\begin{equation}
\left\{
\begin{array}{c}
r_{+}=MG+\sqrt{\left( MG\right) ^{2}-Q^{2}} \\
r_{-}=MG-\sqrt{\left( MG\right) ^{2}-Q^{2}}
\end{array}
\right. ,
\end{equation}
with $f\left( r\right) >0$ for $r>r_{+}$ and $0<r<r_{-}$ . In the wormhole
language, we will say that $r_{-}$ is the inner throat and $r_{+}$ is the
outer throat. In the horizon language $r_{-}$ is a Cauchy horizon and $r_{+}$
is an event horizon. For each root there is a surface gravity defined by
\begin{equation}
\kappa _{\pm }=\lim\limits_{r\rightarrow r_{\pm }}\frac{1}{2}\left|
g_{00}^{\prime }\left( r\right) \right| ,
\end{equation}
whose values are
\begin{equation}
\left\{
\begin{array}{c}
\kappa _{+}=\left( r_{+}-r_{-}\right) /2r_{+}^{2} \\
\kappa _{-}=\left( r_{-}-r_{+}\right) /2r_{-}^{2}
\end{array}
\right.   \label{p12ab}
\end{equation}
and for each surface gravity there exists a bifurcation surface associated
to a wormhole throat. The {\it Hawking temperature} associated with the
surface gravity of the event horizon is
\begin{equation}
T_{H}=\frac{\kappa _{+}}{2\pi }.
\end{equation}
\item[b)]  $MG=Q$. This is the extreme case. The gravitational potential $%
f\left( r\right) $ admits two real coincident solutions located at $%
r_{+}=r_{-}=r_{e}=MG$ and its form is $f\left( r\right) =\left(
1-MG/r\right) ^{2}$. Here we discover that $\kappa _{+}=\kappa _{-}=0$ and $%
T_{H}=0$.
\item[c)]  $MG<Q$. In this case the gravitational potential $f\left(
r\right) $ admits two complex conjugate solutions located at
\begin{equation}
\left\{
\begin{array}{c}
r_{+,i}=MG+i\sqrt{Q^{2}-\left( MG\right) ^{2}} \\
r_{-,i}=MG-i\sqrt{Q^{2}-\left( MG\right) ^{2}}
\end{array}
\right. ,
\end{equation}
respectively.
\end{itemize}
Cases a) and b) imply $Q=0$ when $M=0$. We will restrict our attention on
case a) only.
\section{Quasilocal energy}
\label{p2}
\begin{figure}[bh]
\vbox{\hfil\epsfxsize=4.5cm\epsfbox{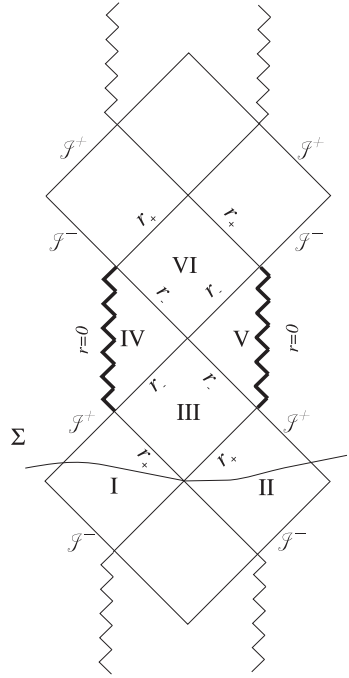}\hfil}
\caption{Penrose diagram for the Reissner-Nordstr\"{o}m
spacetime.} \label{f2}
\end{figure}

Quasilocal energy is defined as the value of the Hamiltonian that
generates unit time translations orthogonal to the two-dimensional boundary%
\footnote{%
See Appendix\ref{app3} for the explicit derivation of the
Hamiltonian in presence of a bifurcation surface.},
\begin{equation}
\Delta E_{Flat}^{RN}\left( M,Q\right) _{|classical}=\frac{1}{8\pi G}%
\int_{S^{2}}d^{2}x\sqrt{\sigma }\left( k-k^{0}\right) ,
\end{equation}
where $\left| N\right| =1$ at $S^{2}$ and $k^{0}$ is the trace of the
extrinsic curvature corresponding to the reference space, which in this case
is flat space. The radial coordinate $x$ continuous on ${\cal M}$ is defined
by
\begin{equation}
dx=\pm \frac{dr}{\sqrt{1-\frac{2MG}{r}+\frac{Q^{2}}{r^{2}}}},  \label{p13}
\end{equation}
which in its integrated form becomes
\begin{equation}
x\left( r\right) =\pm \left( r\sqrt{f\left( r\right) }+MG\ln \left( \frac{%
r-MG+r\sqrt{f\left( r\right) }}{\sqrt{\left( MG\right) ^{2}-Q^{2}}}\right)
\right)
\end{equation}
with $x\left( r\right) =0$ when $r=r_{\pm }$. The surfaces located
at $r_{+}$ and $r_{-}$ are bifurcation surfaces denoted by
$S_{+}^{0}$ and $S_{-}^{0}$, respectively.
\begin{figure}[b]
\vbox{\hfil\epsfxsize=5.5cm\epsfbox{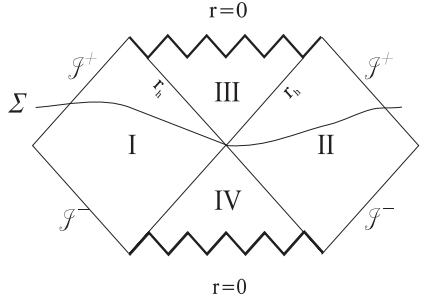}\hfil}
\caption{Penrose diagram for the Schwarzschild spacetime.}
\label{f1}
\end{figure}
In analogy with the Schwarzschild case, see Fig.\ref{f1}, we will
restrict our analysis to the regions I and II of Fig.\ref{f2},
corresponding to the bifurcation surface $S_{+}^{0}$. The constant
time hypersurface we will consider will be denoted by $\Sigma
_{+}=\Sigma _{++}\cup \Sigma _{+-}$, where the plus sign of
Eq.$\left( \ref {p13}\right) $ is relative to $\Sigma _{++}$,
while the minus sign is related to $\Sigma _{+-}$. In $\Sigma
_{++}$ the evaluation of $\Delta E_{Flat}^{RN}\left( M,Q\right)
_{|classical}$ can be obtained as follows: first we consider the
static Einstein-Rosen bridge associated to the RN space
\cite{FroMar,HawHor}
\begin{equation}
ds^{2}=-N^{2}\left( r\right) dt^{2}+g_{xx}dx^{2}+r^{2}\left( x\right)
d\Omega ^{2},  \label{p14}
\end{equation}
where $N$, $g_{xx}$, and $r$ are functions of $x$ defined by Eq.$\left( \ref
{p13}\right) $. Second, we consider the boundary $S_{++}^{2}$, located at $%
x\left( r\right) =\bar{x}^{++}\left( R\right) $, and its associated normal $%
n^{\mu }=\delta _{y}^{\mu }$. The expression of the trace
\begin{equation}
k=-\frac{1}{\sqrt{h}}\left( \sqrt{h}n^{\mu }\right) _{,\mu },
\end{equation}
gives for the RN space
\begin{equation}
k^{RN}=-2\frac{r,_{x}}{r}_{|RN}=-2\frac{\sqrt{f\left( r\right) }}{r}_{|RN}=-%
\frac{2}{r}\sqrt{1-\frac{2MG}{r}+\frac{Q^{2}}{r^{2}}}.
\label{p15}
\end{equation}
Note that if we make the identification $N^{2}=1-\frac{2MG}{r}+\frac{Q^{2}}{%
r^{2}}$, the line element $\left( \ref{p14}\right) $ reduces to the RN
metric written in another form; for our purposes the form of $N\left(
r\right) $ can be left unspecified. Thus the computation of $E_{+}$ gives
\begin{equation}
\Delta E_{Flat}^{RN}\left( M,Q\right) _{|classical}=\frac{1}{8\pi G}%
\int_{S_{++}^{2}}d\Omega ^{2}r^{2}\left[ \frac{-2\sqrt{f\left( r\right) }}{r}%
+\frac{2}{r}\right] _{|r=R}
\end{equation}
\begin{equation}
=-\frac{R}{G}\left[ \sqrt{1-\frac{2MG}{R}+\frac{Q^{2}}{R^{2}}}-1\right] .
\end{equation}
When $R\gg 2MG$, (and therefore $R\gg Q$), we obtain
\begin{equation}
\Delta E_{Flat}^{RN}\left( M,Q\right) _{|classical}\simeq -\frac{R}{G}\left[
\left( 1-\frac{MG}{R}+\frac{Q^{2}}{2R^{2}}\right) -1\right] =M.
\end{equation}
As expected, since the RN space is A.F., we have obtained that the classical
contribution to the energy is exactly the Arnowitt-Deser-Misner mass (ADM)
\cite{Radinschi}. Since the RN metric is A.F., the computation of the
classical energy term leads to
\begin{equation}
E^{RN}\left( M,Q\right) =E^{Flat}+\Delta E_{Flat}^{RN}\left( M,Q\right)
_{|classical}=\Delta E_{Flat}^{RN}\left( M,Q\right) _{|classical},
\end{equation}
which can be computed by means of quasilocal energy. Thus in complete
analogy with the Schwarzschild case, we conclude that flat space cannot
decay into RN space because the associated boundary energy (ADM) is
different, for every value of the mass $M$ included the extreme value.
However if we look at the whole hypersurface $\Sigma _{+}=\Sigma _{++}\cup
\Sigma _{+-}$, the total classical energy becomes
\begin{equation}
E^{RN}\left( M,Q\right) =\Delta E_{Flat}^{RN}\left( M,Q\right)
_{|classical}^{+}+\Delta E_{Flat}^{RN}\left( M,Q\right) _{|classical}^{-}
\end{equation}
with
\[
\Delta E_{Flat}^{RN}\left( M,Q\right) _{|classical}^{+}=\frac{1}{8\pi G}%
\int_{S_{++}^{2}}d^{2}x\sqrt{\sigma }\left( k-k^{0}\right) ,
\]
\begin{equation}
\Delta E_{Flat}^{RN}\left( M,Q\right) _{|classical}^{-}=-\frac{1}{8\pi G}%
\int_{S_{+-}^{2}}d^{2}x\sqrt{\sigma }\left( k-k^{0}\right) .
\label{p16}
\end{equation}
Here the boundaries $S_{++}^{2}$ and $S_{+-}^{2}$ are located in the two
disconnected regions ${\cal M}_{++}$ and ${\cal M}_{+-}$ respectively with
coordinate values $x=\bar{x}^{++}\left( \bar{x}^{+-}\right) $ and the trace
of the extrinsic curvature in both regions is
\begin{equation}
k^{RN}=\left\{
\begin{array}{c}
-2r,_{x}/r\qquad on\ \Sigma _{++} \\
2r,_{x}/r\qquad on\ \Sigma _{+-}
\end{array}
\right. .
\end{equation}
Thus one gets
\begin{equation}
\Delta E_{Flat}^{RN}\left( M,Q\right) _{|classical}^{\pm }=\left\{
\begin{array}{c}
M\qquad on\ S_{++}^{2} \\
-M\qquad on\ S_{+-}^{2}
\end{array}
\right. ,
\end{equation}
where for $E_{-}$ we have used the conventions relative to $\Sigma _{+-}$
and $S_{+-}^{2}$. Therefore for every value of the boundary $R$, (provided
we take symmetric boundary conditions with respect to the bifurcation
surface), we have
\begin{equation}
E^{RN}\left( M,Q\right) =M+\left( -M\right) =0,
\end{equation}
namely the energy is conserved. As stressed in Ref.\cite{FroMar}, since we
have a spacetime with a bifurcation surface, the quantities $\Delta
E_{Flat}^{RN}\left( M,Q\right) _{|classical}^{+}$ and $\Delta
E_{Flat}^{RN}\left( M,Q\right) _{|classical}^{-}$ have the same relative
sign, while the total energy is given by the sum $\Delta E_{Flat}^{RN}\left(
M,Q\right) _{|classical}^{+}+\Delta E_{Flat}^{RN}\left( M,Q\right)
_{|classical}^{-}$\footnote{%
In Ref.\cite{FroMar} we have a subtraction instead of a sum. This
is due to conventions adopted.}. The energies associated to the
boundaries are symmetric and they have the same relative sign
while the total energy reflects the orientation reversal of the
two boundaries. Apparently, Eq.$\left( \ref{p16}\right) $ seems to
violate causality, because of the minus appearance in front of the
expression. However, as reported in Appendix\ref{app3}, it is such
a sign that makes the computation causality preserving. Since the
total classical energy is conserved we can discuss the existence
of an instability. To this aim we refer to the variational
approach to compute the energy density to
one-loop\cite{Remo5,Remo6,Remo7,CJT,CP}.
\section{Energy Density Calculation in Schr\"{o}dinger Representation}
\label{p3}In previous section we have fixed our attention to the
classical part of Eq.$\left( \ref{i0}\right) $. In this section,
we apply the same calculation scheme of
Refs.\cite{Remo5,Remo6,Remo7} to compute one loop corrections to
the classical RN term. Like the Schwarzschild case, there appear
two classical constraints
\begin{equation}
\left\{
\begin{array}{l}
{\cal H}\text{ }=0 \\
{\cal H}^{i}=0
\end{array}
\right.
\end{equation}
and two {\it quantum} constraints
\begin{equation}
\left\{
\begin{array}{l}
{\cal H}\tilde{\Psi}\text{ }=0 \\
{\cal H}^{i}\tilde{\Psi}=0
\end{array}
\right.
\end{equation}
for the Hamiltonian respectively, which are satisfied both by the RN and
flat metric on shell. ${\cal H}\tilde{\Psi}$ $=0$ is known as the {\it %
Wheeler-DeWitt} equation (WDW). Our purpose is the computation of
\begin{equation}
\Delta E_{flat}^{RN}\left( M,Q\right) _{|1-loop}=\frac{\left\langle \Psi
\left| H_{\Sigma }^{RN}-H_{\Sigma }^{flat}\right| \Psi \right\rangle }{%
\left\langle \Psi |\Psi \right\rangle }  \label{p21}
\end{equation}
where $H_{\Sigma }^{RN}$ and $H_{\Sigma }^{flat}$ are the total Hamiltonians
referred to the RN and flat spacetimes respectively for the volume term\cite
{Remo} and $\Psi $ is a wave functional obtained following the usual WKB
expansion of the WDW solution. In this context, the approximated wave
functional will be substituted by a {\it trial wave functional} of the
gaussian form according to the variational approach we shall use to evaluate
$\Delta E_{flat}^{RN}\left( M,Q\right) _{|1-loop}$. To compute such a
quantity we will consider on $\Sigma $ deviations from the RN metric spatial
section of the form,
\begin{equation}
g_{ij}=\bar{g}_{ij}+h_{ij},  \label{p21a}
\end{equation}
where
\begin{equation}
\bar{g}_{ij}dx^{i}dx^{j}=\left( 1-\frac{2MG}{r}+\frac{Q^{2}}{r^{2}}\right)
^{-1}dr^{2}+r^{2}d\Omega ^{2}
\end{equation}
is the spatial RN background. Thus the expansion of the three-scalar
curvature $\int d^{3}x\sqrt{g}R^{\left( 3\right) }$ up to $o\left(
h^{2}\right) $ gives
\[
\int_{\Sigma }d^{3}x\sqrt{\bar{g}}\left[ -\frac{1}{4}h\triangle h+\frac{1}{4}%
h^{li}\triangle h_{li}-\frac{1}{2}h^{ij}\nabla _{l}\nabla _{i}h_{j}^{l}+%
\frac{1}{2}h\nabla _{l}\nabla _{i}h^{li}-\frac{1}{2}h^{ij}R_{ia}h_{j}^{a}+%
\frac{1}{2}hR_{ij}h^{ij}\right]
\]
\begin{equation}
+\int_{\Sigma }d^{3}x\sqrt{\bar{g}}\left[ \frac{1}{4}h\left( R^{\left(
0\right) }\right) h\right] ,
\end{equation}
where $R^{\left( 0\right) }$ is the three-scalar curvature on-shell. To
explicitly make calculations, we need an orthogonal decomposition for both $%
\pi _{ij\text{ }}$and $h_{ij}$ to disentangle gauge modes from physical
deformations. We define the inner product
\begin{equation}
\left\langle h,k\right\rangle :=\int_{\Sigma }\sqrt{g}G^{ijkl}h_{ij}\left(
x\right) k_{kl}\left( x\right) d^{3}x,
\end{equation}
by means of the inverse WDW metric $G_{ijkl}$, to have a metric on the space
of deformations, i.e. a quadratic form on the tangent space at h, with
\begin{equation}
G^{ijkl}=(g^{ik}g^{jl}+g^{il}g^{jk}-2g^{ij}g^{kl}).
\end{equation}
The inverse metric is defined on co-tangent space and it assumes the form
\begin{equation}
\left\langle p,q\right\rangle :=\int_{\Sigma }\sqrt{g}G_{ijkl}p^{ij}\left(
x\right) q^{kl}\left( x\right) d^{3}x\text{,}
\end{equation}
so that
\begin{equation}
G^{ijnm}G_{nmkl}=\frac{1}{2}\left( \delta _{k}^{i}\delta _{l}^{j}+\delta
_{l}^{i}\delta _{k}^{j}\right) .
\end{equation}
Note that in this scheme the ``inverse metric'' is actually the WDW metric
defined on phase space. Now, we have the desired decomposition on the
tangent space of 3-metric deformations\cite{BerEbi,York}:
\begin{equation}
h_{ij}=\frac{1}{3}hg_{ij}+\left( L\xi \right) _{ij}+h_{ij}^{\bot }
\end{equation}
where the operator $L$ maps $\xi _{i}$ into symmetric tracefree tensors
\begin{equation}
\left( L\xi \right) _{ij}=\nabla _{i}\xi _{j}+\nabla _{j}\xi _{i}-\frac{2}{3}%
g_{ij}\left( \nabla \cdot \xi \right) .
\end{equation}
Then the inner product between three-geometries becomes
\[
\left\langle h,h\right\rangle :=\int_{\Sigma }\sqrt{g}G^{ijkl}h_{ij}\left(
x\right) h_{kl}\left( x\right) d^{3}x=
\]
\begin{equation}
\int_{\Sigma }\sqrt{g}\left[ -\frac{2}{3}h^{2}+\left( L\xi \right)
^{ij}\left( L\xi \right) _{ij}+h^{ij\bot }h_{ij}^{\bot }\right] .
\end{equation}
With the orthogonal decomposition in hand we can define the trial wave
functional
\begin{equation}
\Psi \left[ h_{ij}\left( \overrightarrow{x}\right) \right] ={\cal N}\exp
\left\{ -\frac{1}{4}\left[ \left\langle hK^{-1}h\right\rangle _{x,y}^{\bot
}+\left\langle \left( L\xi \right) K^{-1}\left( L\xi \right) \right\rangle
_{x,y}^{\Vert }+\left\langle hK^{-1}h\right\rangle _{x,y}^{Trace}\right]
\right\} ,
\end{equation}
where ${\cal N}$ is a normalization factor. Since we are only interested in
the perturbations of the physical degrees of freedom, we will only fix our
attention on the TT (traceless and transverseless) tensor sector, therefore
reducing the previous form into\footnote{%
In this paper we have defined the trial wave functional without a Planck
length constant factor in the exponent. This choice alters momentarily the
physical dimensions of the problem which are reestablished after the
variational procedure.}
\begin{equation}
\Psi \left[ h_{ij}\left( \overrightarrow{x}\right) \right] ={\cal N}\exp
\left\{ -\frac{1}{4}\left\langle hK^{-1}h\right\rangle _{x,y}^{\bot
}\right\} .
\end{equation}
This restriction is motivated by the fact that if an instability appears
this will be in the physical sector referred to TT tensors, namely a
nonconformal instability. This choice seems to be corroborated by the action
decomposition of\cite{GriKos}, where only TT tensors contribute to the
partition function\footnote{%
See also\cite{EKP} for another point of view.}. To calculate the energy
density, we need to know the action of some basic operators on $\Psi \left[
h_{ij}\right] $\cite{CJT}. The action of the operator $h_{ij}$ on $|\Psi
\rangle =\Psi \left[ h_{ij}\right] $ is realized by
\begin{equation}
h_{ij}\left( x\right) |\Psi \rangle =h_{ij}\left( \overrightarrow{x}\right)
\Psi \left[ h_{ij}\right] ,
\end{equation}
while the action of the operator $\pi _{ij}$ on $|\Psi \rangle $, in
general, is
\begin{equation}
\pi _{ij}\left( x\right) |\Psi \rangle =-i\frac{\delta }{\delta h_{ij}\left(
\overrightarrow{x}\right) }\Psi \left[ h_{ij}\right] .
\end{equation}
The inner product is defined by the functional integration
\begin{equation}
\left\langle \Psi _{1}\mid \Psi _{2}\right\rangle =\int \left[ {\cal D}h_{ij}%
\right] \Psi _{1}^{\ast }\left\{ h_{ij}\right\} \Psi _{2}\left\{
h_{kl}\right\}
\end{equation}
and by applying previous functional integration rules, we obtain the
expression of the one-loop-like Hamiltonian form for TT (traceless and
transverseless) deformations\cite{Remo5,Remo6,Remo7}
\begin{equation}
H^{\bot }=\frac{1}{4}\int_{{\cal M}}d^{3}x\sqrt{g}G^{ijkl}\left[ \left(
16\pi G\right) K^{-1\bot }\left( x,x\right) _{ijkl}+\frac{1}{16\pi G}\left(
\triangle _{2}\right) _{j}^{a}K^{\bot }\left( x,x\right) _{iakl}\right] .
\label{p22}
\end{equation}
The propagator $K^{\bot }\left( x,x\right) _{iakl}$ comes from a functional
integration and it can be represented as
\begin{equation}
K^{\bot }\left( \overrightarrow{x},\overrightarrow{y}\right)
_{iakl}:=\sum_{\tau}\frac{h_{ia}^{\bot }\left(
\overrightarrow{x}\right) h_{kl}^{\bot }\left(
\overrightarrow{y}\right) }{2\lambda\left(\tau\right) },
\end{equation}
where $h_{ia}^{\bot }\left( \overrightarrow{x}\right) $ are the
eigenfunctions of $\triangle _{2}$. $\tau$ denotes a complete set
of indices and $\lambda\left(\tau\right) $ are a set of
variational parameters to be determined by the minimization of
Eq.$\left( \ref{p22}\right) $.
\section{The Reissner-Nordstr\"{o}m Metric spin 2 operator and the
evaluation of the energy density}
\label{p4}To evaluate the energy density, we are led to study the following
eigenvalue equation
\begin{equation}
\left( \triangle _{2}\right) _{j}^{a}h_{a}^{i}=E^{2}h_{j}^{i}  \label{p41}
\end{equation}
where $\left( \triangle _{2}\right) _{j}^{a}$ is the Spin-two operator for
the RN metric defined by
\begin{equation}
\left( \triangle _{2}\right) _{j}^{a}:=-\triangle \delta _{j}^{a}+2V_{j}^{a}.
\label{p42}
\end{equation}
$\triangle $ is the curved Laplacian (Laplace-Beltrami operator) on a RN
background and $V_{j\text{ }}^{a}$ is defined as
\begin{equation}
V_{j}^{a}:=R_{j}^{a}-P_{j}^{a},  \label{p43}
\end{equation}
where $R_{j}^{a}$ is the mixed Ricci tensor whose components are:
\begin{equation}
R_{i}^{a}=\left\{ -\frac{2MG}{r^{3}}+\frac{2Q^{2}}{r^{4}},\frac{MG}{r^{3}},%
\frac{MG}{r^{3}}\right\}
\end{equation}
and $P_{j}^{a}$ is a mixed tensor coming from the electromagnetic
energy-momentum tensor expanded to second order in $h_{ij}$\footnote{%
See Appendix \ref{app2}.}. We will follow Regge and Wheeler in analyzing the
equation as modes of definite frequency, angular momentum and parity\cite{RW}%
. The quantum number corresponding to the projection of the angular momentum
on the z-axis will be set to zero. This choice will not alter the
contribution to the total energy since we are dealing with a spherical
symmetric problem. In this case, Regge-Wheeler decomposition shows that the
even-parity three-dimensional perturbation is $h_{ij}^{even}\left(
r,\vartheta ,\phi \right) =$%
\begin{equation}
diag\left[ H\left( r\right) \left( 1-\frac{2MG}{r}+\frac{Q^{2}}{r^{2}}%
\right) ^{-1},r^{2}K\left( r\right) ,r^{2}\sin ^{2}\vartheta K\left(
r\right) \right] Y_{l0}\left( \vartheta ,\phi \right) .  \label{p44}
\end{equation}
In this representation $H\left( r\right) $ and $K\left( r\right) $ behave as
they were scalar fields and the Laplacian restricted to $\Sigma $ is
\begin{equation}
\triangle _{l}=\left( 1-\frac{2MG}{r}+\frac{Q^{2}}{r^{2}}\right) \frac{d^{2}%
}{dr^{2}}+\left( \frac{2r-3MG}{r^{2}}+\frac{Q^{2}}{r^{3}}\right) \frac{d}{dr}%
-\frac{l\left( l+1\right) }{r^{2}}.  \label{p45}
\end{equation}
The mixed tensor in Eq. $\left( \ref{p43}\right) $ assumes a different form
according to whether we are dealing with the magnetic or electric charge.
\subsection{The electric charge contribution}
\label{ep4}
\begin{figure}[tbh]
\vbox{\hfil\epsfxsize=4.5cm\epsfbox{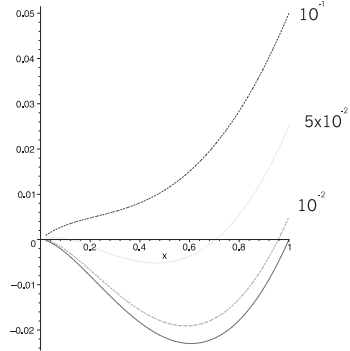}\hfil} \caption{Plot
of Electric charge contribution for different values of the
parameter $\alpha _{e}^{2}$. For $\alpha _{e}^{2}=0$, we have the
 Schwarzschild space.} \label{f3}
\end{figure}

In this case the mixed tensor $P_{j}^{a}$ becomes
\begin{equation}
P_{j}^{a}:=\left( T_{\alpha \beta }u^{\alpha }u^{\beta }\right) \delta
_{j}^{a}-\frac{\kappa }{2\pi }\left( F_{01}\right) ^{2}\delta _{1}^{a}\delta
_{j}^{1}
\end{equation}
and for a generic value of the angular momentum $L$, the system $\left( \ref
{p41}\right) $ becomes
\begin{equation}
\left\{
\begin{array}{c}
\left( -\triangle _{l}-\frac{4MG}{r^{3}}+\frac{10Q_{e}^{2}}{r^{4}}\right)
H_{l}\left( r\right) =E_{l,H}^{2}H_{l}\left( r\right)  \\
\\
\left( -\triangle _{l}+\frac{2MG}{r^{3}}-\frac{2Q_{e}^{2}}{r^{4}}\right)
K_{l}\left( r\right) =E_{l,K}^{2}K_{l}\left( r\right)
\end{array}
,\right.   \label{ep41}
\end{equation}
where $E_{l,H}^{2}$ and $E_{l,K}^{2}$ are the eigenvalues for the $%
H_{l}\left( r\right) $ field and the $K_{l}\left( r\right) $ field
respectively. Defining reduced fields
\begin{equation}
H_{l}\left( r\right) =\frac{h_{l}\left( r\right) }{r};\qquad K_{l}\left(
r\right) =\frac{k_{l}\left( r\right) }{r},
\end{equation}
and passing to the proper geodesic distance from the {\it throat} of the
bridge defined by Eq.$\left( \ref{p13}\right) $, the system $\left( \ref
{ep41}\right) $ becomes\footnote{%
The system is invariant in form if we make the minus choice in Eq.$\left(
\ref{p13}\right) $.}
\begin{equation}
\left\{
\begin{array}{c}
-\frac{d^{2}}{dx^{2}}h_{l}\left( x\right) +V_{l}^{+}\left( x\right)
h_{l}\left( x\right) =E_{l}^{2}h\left( x\right)  \\
\\
-\frac{d^{2}}{dx^{2}}k_{l}\left( x\right) +V_{l}^{-}\left( x\right)
k_{l}\left( x\right) =E_{l}^{2}k\left( x\right) ,
\end{array}
\right.   \label{ep42}
\end{equation}
where
\begin{equation}
V_{l}^{\mp }\left( x\right) =\frac{l\left( l+1\right) }{r^{2}\left( x\right)
}\mp \frac{3MG}{r^{3}\left( x\right) }\pm \frac{6Q_{e}^{2}}{r^{4}\left(
x\right) }+\frac{3Q_{e}^{2}}{r^{4}\left( x\right) }.
\end{equation}
When $r\gg r_{+}$ $x\left( r\right) \simeq r$, then we can approximate the
potential with
\begin{equation}
V_{l}^{\mp }\left( x\right) \simeq \frac{l\left( l+1\right) }{x^{2}}
\end{equation}
and the solution for the system $\left( \ref{ep42}\right) $ is
\begin{equation}
h_{pl}\left( x\right) =k_{pl}\left( x\right) =\sqrt{\frac{2}{\pi }}\left(
px\right) j_{l}\left( px\right) ,  \label{ep42aa}
\end{equation}
where $j_{l}\left( px\right) $ is the spherical Bessel function. If we
consider flat space, i.e. $M=Q=0$, the system $\left( \ref{ep42}\right) $
becomes
\begin{equation}
\left\{
\begin{array}{c}
\left( -\frac{d^{2}}{dr^{2}}+\frac{l\left( l+1\right) }{r^{2}}\right)
h_{l}\left( r\right) =E_{l}^{2}h_{l}\left( r\right)  \\
\\
\left( -\frac{d^{2}}{dr^{2}}+\frac{l\left( l+1\right) }{r^{2}}\right)
k_{l}\left( r\right) =E_{l}^{2}k_{l}\left( r\right)
\end{array}
\right. ,
\end{equation}
and the solution is
\begin{equation}
h_{pl}\left( r\right) =k_{pl}\left( r\right) =\sqrt{\frac{2}{\pi }}\left(
pr\right) j_{l}\left( pr\right) .
\end{equation}
On the other hand when $r\longrightarrow r_{0}>r_{+}$%
\begin{equation}
x\left( r\right) \simeq \sqrt{2\kappa _{+}\left( r-r_{+}\right) }\qquad
V_{l}^{\mp }\left( x\right) \longrightarrow \frac{l\left( l+1\right) }{%
r_{0}^{2}}\mp \frac{3MG}{r_{0}^{3}}\pm \frac{6Q_{e}^{2}}{r_{0}^{4}}+\frac{%
3Q_{e}^{2}}{r_{0}^{4}}=const,
\end{equation}
where $\kappa _{+}$ has been defined in Eq. $\left( \ref{p12ab}\right) $.
However to use the simplicity of Bessel functions for flat and curved space
when $r\longrightarrow r_{0}>r_{+}$, we approximate the potential with
\begin{equation}
V_{l}^{\mp }\left( x\right) =\frac{l\left( l+1\right) }{x^{2}}\mp \frac{3MG}{%
r_{0}^{3}}\pm \frac{6Q_{e}^{2}}{r_{0}^{4}}+\frac{3Q_{e}^{2}}{r_{0}^{4}}.
\end{equation}
Thus close to the wormhole throat we experience a potential barrier
(potential hole) whose solution, for system $\left( \ref{ep42}\right) $ is
\begin{equation}
\left\{
\begin{array}{cc}
h_{P_{-}l}\left( x\right) = & \sqrt{\frac{2}{\pi }}\left( P_{-}x\right)
j_{l}\left( P_{-}x\right)  \\
k_{P_{+}l}\left( x\right) = & \sqrt{\frac{2}{\pi }}\left( P_{+}x\right)
j_{l}\left( P_{+}x\right)
\end{array}
\right. .  \label{ep42ab}
\end{equation}
$P_{\mp }$ is such that $P_{\mp }^{2}=E_{l}^{2}\mp c_{\mp }^{2}$, where%
\footnote{%
Actually a more standard approach of this problem can be considered by means
of phase shifts $\delta _{l}\left( p\right) $.}%
\footnote{%
This choice is also dictated by the necessity of avoiding that different
values of the angular momentum enter in the approximate potential $%
V_{l}^{\pm }\left( x\right) $ couple like in Refs.\cite{Remo,Remo5,Remo6}.}
\begin{equation}
c_{\mp }^{2}=\mp \frac{3MG}{r_{0}^{3}}\pm \frac{6Q_{e}^{2}}{r_{0}^{4}}+\frac{%
3Q_{e}^{2}}{r_{0}^{4}}.  \label{ep42ac}
\end{equation}
Note that for functions described by Eq.$\left( \ref{ep42aa}\right) $ or Eq.$%
\left( \ref{ep42ab}\right) $ we have that
\begin{equation}
h_{l}\left( x\right) ,k_{l}\left( x\right) \rightarrow 0\text{\qquad
when\qquad }x\left( r\right) \rightarrow x\left( r_{+}\right) \simeq 0.
\end{equation}
Thus the propagator becomes
\begin{equation}
K_{\pm }^{\bot }\left( x,y\right) =\frac{1}{r\left(
x\right)r\left( y\right)}\sum_{P,l,m}\left\{
\begin{array}{c}
k_{P_{+}l}\left( x\right) k_{P_{+}l}\left( y\right) /\lambda _{+}\left(
P\right)  \\
h_{P_{-}l}\left( x\right) h_{P_{-}l}\left( y\right) /\lambda _{-}\left(
P\right)
\end{array}
\right. Y_{l0}\left( \vartheta ,\phi \right) Y_{l0}\left(
\vartheta^{\prime } ,\phi^{\prime } \right)   \label{ep42a}
\end{equation}
$\lambda _{\pm }\left( P\right) $ is referred to the potential function $%
V_{l}^{\pm }\left( x\right) $. This is the more general expression for the
propagator. Indeed, when one considers flat space or the region far away
from the throat $r_{+}$, it is sufficient the substitution of $P_{\mp }$
with $p$. Inserting Eq.$\left( \ref{ep42a}\right) $ into Eq.$\left( \ref{p22}%
\right) $ one gets
\begin{equation}
E\left( M,Q_{e},\lambda \right) =\frac{1}{2\pi }\sum_{p,l}\frac{\left(
2l+1\right) }{4\pi }\sum_{i=1}^{2}\left[ \left( 16\pi G\right) \lambda
_{i}\left( p\right) +\frac{E_{i}^{2}\left( p,M,Q_{e}\right) }{\left( 16\pi
G\right) \lambda _{i}\left( p\right) }\right]   \label{ep42b}
\end{equation}
where
\begin{equation}
\left\{
\begin{array}{c}
E_{1}^{2}\left( p,M,Q_{e}\right) =p^{2}-\frac{3MG}{r_{0}^{3}}+\frac{%
9Q_{e}^{2}}{r_{0}^{4}} \\
E_{2}^{2}\left( p,M,Q_{e}\right) =p^{2}+\frac{3MG}{r_{0}^{3}}-\frac{%
3Q_{e}^{2}}{r_{0}^{4}}
\end{array}
\right. ,
\end{equation}
$\lambda _{i}\left( p\right) $ are variational parameters corresponding to
the eigenvalues for a (graviton) spin-two particle in an external field. By
minimizing $\left( \ref{ep42b}\right) $ with respect to $\lambda _{i}\left(
p\right) $ one obtains $\overline{\lambda }_{i}\left( p\right) =\left[
E_{i}^{2}\left( p,M,Q_{e}\right) \right] ^{\frac{1}{2}}/\left( 16\pi
G\right) $ and
\begin{equation}
E\left( M,Q_{e},\overline{\lambda }\right) =\frac{1}{\pi }\sum_{p,l}\frac{%
\left( 2l+1\right) }{4\pi }\left[ \sqrt{E_{1}^{2}\left( p,M,Q_{e}\right) }+%
\sqrt{E_{2}^{2}\left( p,M,Q_{e}\right) }\right]
\end{equation}
with
\begin{equation}
p^{2}+c_{\mp }^{2}>0,
\end{equation}
where we have used definition $\left( \ref{ep42ac}\right) $. We can evaluate
the total energy for the electric RN background by replacing the sum with an
integral leading to
\begin{equation}
E\left( M,Q_{e}\right) =\frac{V}{4\pi ^{2}}\int_{0}^{\infty }dpp^{2}\left(
\sqrt{p^{2}+c_{-}^{2}}+\sqrt{p^{2}+c_{+}^{2}}\right) ,  \label{ep42c}
\end{equation}
where $V$ is the volume localized near the wormhole throat. For
flat space we put $M=Q=0$ and we get
\begin{equation}
E\left( 0\right) =\frac{V}{2\pi ^{2}}\int_{0}^{\infty }dpp^{3}  \label{ep42d}
\end{equation}
Now, we are in position to compute the difference between $\left( \ref{ep42c}%
\right) $ and $\left( \ref{ep42d}\right) $. The explicit evaluation of the
integrals of Eq.$\left( \ref{ep42c}\right) $ in the U.V. limit, gives
\begin{equation}
E\left( M,Q_{e}\right) \simeq \frac{V}{4\pi ^{2}}\frac{1}{4}\left[
2p^{4}+p^{2}\left( c_{+}^{2}+c_{-}^{2}\right) -\frac{c_{+}^{4}}{4}\ln \left(
\frac{p^{2}}{c_{+}^{2}}\right) -\frac{c_{-}^{4}}{4}\ln \left( \frac{p^{2}}{%
c_{-}^{2}}\right) \right] .
\end{equation}
Thus
\[
\Delta E_{flat}^{RN}\left( M,Q_{e}\right) =E\left( M,Q_{e}\right) -E\left(
0\right)
\]
\[
=\frac{V}{4\pi ^{2}}\int_{0}^{\infty }dpp^{2}\left[ \sqrt{p^{2}+c_{-}^{2}}+%
\sqrt{p^{2}+c_{+}^{2}}-2\sqrt{p^{2}}\right]
\]
\[
\simeq \frac{V}{4\pi ^{2}}\left[ \frac{1}{4}p^{2}\left(
c_{+}^{2}+c_{-}^{2}\right) -\frac{c_{+}^{4}}{16}\ln \left( \frac{p^{2}}{%
c_{+}^{2}}\right) -\frac{c_{-}^{4}}{16}\ln \left( \frac{p^{2}}{c_{-}^{2}}%
\right) \right]
\]
\begin{equation}
\simeq \frac{V}{4\pi ^{2}}\left[ \frac{1}{4}p^{2}\left(
c_{+}^{2}+c_{-}^{2}\right) -\frac{\left( c_{-}^{4}+c_{+}^{4}\right) }{16}\ln
\left( \frac{\Lambda ^{2}}{\mu ^{2}}\right) \right] ,  \label{ep43}
\end{equation}
where we have used the approximation $p^{2}>>c_{\mp }^{2}$ and a cut-off $%
\Lambda \leq m_{p}$ to keep under control the $UV$ divergence and we have
introduced an arbitrary scale $\mu $. In particular for the Schwarzschild
case, $\mu $ has been determined by the quantity $3MG/r_{0}^{3}=c_{M}^{2}$
and in that case the approximated expression for $\Delta E\left( M\right) $
is
\begin{equation}
\Delta E\left( M\right) \sim -\frac{V}{2\pi ^{2}}\frac{c_{M}^{4}}{16}\ln
\left( \frac{\Lambda ^{2}}{c_{M}^{2}}\right) =-\frac{V}{32\pi ^{2}}\left(
\frac{3MG}{r_{0}^{3}}\right) ^{2}\ln \left( \frac{r_{0}^{3}\Lambda ^{2}}{3MG}%
\right) .
\end{equation}
Nevertheless, in this paper we would like to establish if a mechanism of
space-time foam formation can arise in competition with the foam model
created by Schwarzschild wormholes. To this purpose in this case we shall
fix as a scale the value $c_{M}^{2}$ and we define the dimensionless
parameter $\alpha =\left( Q_{e}/MG\right) $. Thus Eq. $\left( \ref{ep43}%
\right) $ becomes
\begin{equation}
\Delta E_{flat}^{RN}\left( M,Q_{e}\right) \simeq \frac{V}{4\pi ^{2}}\Lambda
^{4}\left[ \frac{1}{2}x\alpha _{e}^{2}+\frac{1}{16}\left( 1+5\alpha
_{e}^{4}-4\alpha _{e}^{2}\right) x^{2}\ln x\right]   \label{ep43a}
\end{equation}
where we have defined
\begin{equation}
\alpha _{e}^{2}=\alpha ^{2}\frac{MG}{r_{0}};\quad x=\frac{3MG}{%
r_{0}^{3}\Lambda ^{2}}.
\end{equation}
Since $MG>Q$ and since $r_{0}>MG>Q$ then $0<\alpha _{e}^{2}<1$. In
Fig.\ref {f3}, we show the plot of Eq.$\left( \ref{ep43a}\right) $
for different values of the parameter $\alpha _{e}^{2}$.
Since only the minimum of $\Delta E_{flat}^{RN}\left(
M,Q_{e}\right) $ is physically relevant, we see that for $\alpha
_{e}^{2}=10^{-1}$, the minimum is positive. This means that the
electric field tends to dominate the gravitational attraction and
becomes more and more stronger as $\alpha _{e}^{2}$ approaches the
extreme value, i.e. $MG=Q_{e}$. As expected for small values of
$\alpha _{e}^{2}$, the behavior for the energy gap of the
Schwarzschild and RN backgrounds with flat space as reference
space are very close. Nevertheless, the energy gap of the last one
is always higher than the same one computed with the Schwarzschild
background. This means that even for very small electric charges,
the Casimir energy for RN wormholes is always higher than the
Casimir energy for Schwarzschild wormholes.
\subsection{The magnetic charge contribution}
\label{mp4}
\begin{figure}[tbh]
\vbox{\hfil\epsfxsize=4.5cm\epsfbox{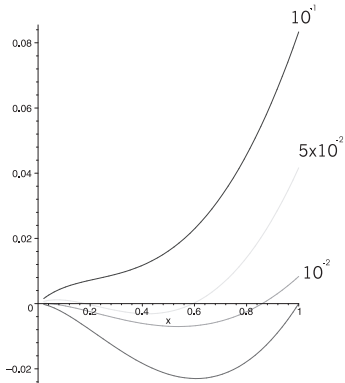}\hfil} \caption{Plot
of Magnetic charge contribution for different values of the
parameter $\alpha _{m}^{2}$. For $\alpha _{m}^{2}=0$, we have the
Schwarzschild space.} \label{f4}
\end{figure}

For the magnetic case $P_{j\text{ }}^{a}$ is
\begin{equation}
P_{j}^{a}:=\kappa \left( T_{\alpha \beta }u^{\alpha }u^{\beta }\right)
\delta _{j}^{a}-2\left( F_{23}\right) ^{2}\left( \delta _{2}^{a}\delta
_{j}^{2}+\delta _{3}^{a}\delta _{j}^{3}+\delta _{2}^{a}\delta
_{j}^{3}\right) .
\end{equation}
Since the last term in previous equation mixes the components $h_{2}^{2}$
and $h_{3}^{3}$, we have
\begin{equation}
\left[ \left( -\triangle _{l}+\frac{2MG}{r^{3}}+\frac{6Q_{m}^{2}}{r^{4}}%
\right) {\bf I}_{2}+\frac{4Q_{m}^{2}}{r^{4}}\left(
\begin{array}{cc}
0 & 1 \\
1 & 0
\end{array}
\right) \right] \left(
\begin{array}{c}
h_{2}^{2} \\
h_{3}^{3}
\end{array}
\right) .  \label{mp40}
\end{equation}
By repeating the same steps of section \ref{ep4}, the system $\left( \ref
{p41}\right) $ becomes
\begin{equation}
\left\{
\begin{array}{c}
\left( -\triangle _{l}-\frac{4MG}{r^{3}}+\frac{2Q_{m}^{2}}{r^{4}}\right)
H\left( r\right) =E_{l,H}^{2}H\left( r\right)  \\
\\
\left( -\triangle _{l}+\frac{2MG}{r^{3}}+\frac{10Q_{m}^{2}}{r^{4}}\right)
K\left( r\right) =E_{l,K}^{2}K\left( r\right)
\end{array}
\right. ,  \label{mp41}
\end{equation}
where we have diagonalized the operator in $\left( \ref{mp40}\right) $.
Since the diagonalization gives the eigenvalues $\pm 1$ with eigenvectors $%
\frac{h_{2}^{2}\pm h_{3}^{3}}{\sqrt{2}}$ respectively, we have considered
only the positive eigenvalue since the negative one has the eigenvector $%
\frac{h_{2}^{2}-h_{3}^{3}}{\sqrt{2}}$ which vanishes because $%
h_{2}^{2}=h_{3}^{3}$ in Regge-Wheeler representation. Thus following the
steps we have used for the electric part, we arrive to
\begin{equation}
\left\{
\begin{array}{c}
-\frac{d^{2}}{dx^{2}}h\left( x\right) +V_{l}^{+}\left( x\right) h\left(
x\right) =E_{l}^{2}h\left( x\right)  \\
\\
-\frac{d^{2}}{dx^{2}}k\left( x\right) +V_{l}^{-}\left( x\right) k\left(
x\right) =E_{l}^{2}k\left( x\right)
\end{array}
,\right.   \label{mp42}
\end{equation}
where
\begin{equation}
V_{l}^{\mp }\left( x\right) =\frac{l\left( l+1\right) }{r^{2}\left( x\right)
}\mp \frac{3MG}{r\left( x\right) ^{3}}\mp \frac{4Q_{m}^{2}}{r\left( x\right)
^{4}}+\frac{5Q_{m}^{2}}{r\left( x\right) ^{4}}.
\end{equation}
In analogy with the electric case, we obtain the approximate solutions of
system $\left( \ref{mp42}\right) $, by restricting the analysis to the
sector where $r\longrightarrow r_{0}>r_{+}.$Then
\begin{equation}
\left\{
\begin{array}{cc}
h_{P_{-}l}\left( x\right) = & \sqrt{\frac{2}{\pi }}\left( P_{-}x\right)
j_{l}\left( P_{-}x\right)  \\
k_{P_{+}l}\left( x\right) = & \sqrt{\frac{2}{\pi }}\left( P_{+}x\right)
j_{l}\left( P_{+}x\right)
\end{array}
\right. .
\end{equation}
$P_{\mp }$ is such that $P_{\mp }^{2}=E_{l}^{2}\mp d_{\mp }^{2}$, where
\begin{equation}
d_{\mp }^{2}=\mp \frac{3MG}{r_{0}^{3}}\mp \frac{4Q_{m}^{2}}{r_{0}^{4}}+\frac{%
5Q_{m}^{2}}{r_{0}^{4}}.  \label{mp42a}
\end{equation}
Finally we arrive to
\begin{equation}
E\left( M,Q_{m},\overline{\lambda }\right) =\frac{V}{4\pi ^{2}}%
\sum_{i=1}^{2}\int_{0}^{\infty }dpp^{2}\sqrt{E_{i}^{2}\left(
p,M,Q_{m}\right) }\text{ }
\end{equation}
with
\begin{equation}
\left\{
\begin{array}{c}
E_{1}^{2}\left( p,M,Q_{m}\right) =p^{2}-\frac{3MG}{r_{0}^{3}}+\frac{Q_{m}^{2}%
}{r_{0}^{4}} \\
E_{2}^{2}\left( p,M,Q_{m}\right) =p^{2}+\frac{3MG}{r_{0}^{3}}+\frac{%
9Q_{m}^{2}}{r_{0}^{4}}
\end{array}
\right. ,
\end{equation}
with the usual condition
\begin{equation}
p^{2}+d_{\mp }^{2}>0,
\end{equation}
where we have used definition $\left( \ref{mp42a}\right) $. For the magnetic
RN background we get
\begin{equation}
E\left( M,Q_{m}\right) =\frac{V}{4\pi ^{2}}\int_{0}^{\infty }dpp^{2}\left(
\sqrt{p^{2}+d_{-}^{2}}+\sqrt{p^{2}+d_{+}^{2}}\right) .
\end{equation}
The zero point energy for flat space is given by Eq. $\left( \ref{ep42d}%
\right) $, then the Casimir energy is given by Eq. $\left( \ref{ep43}\right)
$. Proceeding like the electric case, we find
\begin{equation}
\Delta E_{flat}^{RN}\left( M,Q_{m}\right) \simeq \frac{V}{4\pi ^{2}}\Lambda
^{4}\left[ \frac{5}{6}x\alpha _{m}^{2}+\frac{1}{8}\left( 1+\frac{41}{9}%
\alpha _{m}^{4}+\frac{4}{3}\alpha _{m}^{2}\right) x^{2}\ln
x\right] \label{mp42b}
\end{equation}
where we have defined $\alpha _{m}^{2}=\alpha ^{2}MG/r_{0}$ and
$x$ is the scale variable we have used for the electric case and
$0<\alpha _{m}^{2}<1$. In Fig.\ref{f4}, we show the plot of
Eq.$\left( \ref{mp42b}\right) $ for different values of the
parameter $\alpha _{m}^{2}$
Even for the magnetic case, we find a behavior of the energy gap
similar to the electric one. This means that even in this case,
the Casimir energy for RN wormholes is always higher than the
Casimir energy for Schwarzschild wormholes.
\section{Summary and Conclusions}
\label{p5}In this paper we have considered the possibility of
forming a foamy spacetime using a collection of $N$ coherent RN
wormholes. By applying the same methods used for the Schwarzschild
wormholes, we have found that in case of a electric or magnetic
potential the Casimir energy is always higher than the one
computed for the Schwarzschild case, i.e.
\begin{equation}
\Delta E_{flat}^{RN}\left( M,Q\right) > \Delta
E_{flat}^{Schwarzschild}\left( M\right)
\end{equation}. To this purpose to correctly compare such
energies the same reference scale has been considered. We would
like to remark that for very small electromagnetic field energy
 contribution, it is the total energy difference between the RN spacetime and flat space
 which is negative. The single energy contribution of flat space and RN space
 is strictly positive. Note that this result has been obtained
 when the number of wormholes is equal to one. Indeed the complete
conclusion can be supported when one proofs:
\begin{description}
\item[a)]  the existence of an unstable mode;
\item[b)]  a boundary reduction mechanism coming into play in such a way to
stabilize the system.
\end{description}
Nevertheless, the purpose of the paper is to proof that the Casimir energy
for RN wormholes is always higher than the Schwarzschild Casimir energy.
Since this is the case, it is sufficient to assume the validity of point a)
and b) to conclude\footnote{%
For the point a), the existence of an instability has been proved in Ref.
\cite{Weinberg} only for RN solutions with magnetic charge. As far as we
know the case with electric charge seems to be stable. Nevertheless, it is
interesting to note that the AdS$_{4}$-RN shows such an instability\cite
{SSGIM}}. Note that the result is obtained for every value of $\alpha
_{e}^{2}$ and $\alpha _{m}^{2}$ s.t. $0<
\begin{array}{c}
\alpha _{e}^{2} \\
\alpha _{m}^{2}
\end{array}
<1$. This seems to suggest that a space-time foam formation
realized by RN wormholes is suppressed when compared with the
foamy space formed by Schwarzschild wormholes. On the other hand
one can think to the collection of $N$ RN wormholes as an excited
state with respect to the collection of $N$ Schwarzschild
wormholes leading to the conclusion that such a collection can be
considered as a good candidate for a possible ground state of a
quantum theory of the gravitational field, when compared to a
superposition of large $N$ RN wormholes.
\section{Acknowledgments}
I wish to thank J. D. Bekenstein, R. Brout, M. Cavagli\`{a}, V.
Frolov, C. Kiefer, P. Spindel and C. Vaz for useful comments and
discussions.
\appendix
\section{Kruskal-Szekeres coordinates for RN spacetime}
\label{app1}
We have defined the RN line element in Eq.$\left( \ref{p11}\right)
$. To introduce the Kruskal-Szekeres\cite{KS,HawEll,MTW} type
coordinates we consider the following transformation
\[
ds^{2}=-\left( 1-\frac{2MG}{r}+\frac{Q^{2}}{r^{2}}\right) \left[
dt^{2}-dr^{\ast 2}\right] +r^{2}d\Omega ^{2}
\]
\begin{equation}
=-\left( 1-\frac{2MG}{r}+\frac{Q^{2}}{r^{2}}\right) dvdu+r^{2}\left(
u,v\right) d\Omega ^{2},  \label{a1}
\end{equation}
where $v=t+r^{\ast }$ is the ingoing radial null coordinate and $u=t-r^{\ast
}$ is the outgoing radial null coordinate. The ``tortoise coordinate'' $%
r^{\ast }$ is defined by
\begin{equation}
dr^{\ast }=\frac{r^{2}dr}{\left( r-r_{-}\right) \left( r-r_{+}\right) }
\end{equation}
\begin{equation}
\Longrightarrow r^{\ast }=r+\frac{1}{2\kappa _{+}}\ln \frac{\left|
r-r_{+}\right| }{r_{+}}+\frac{1}{2\kappa _{-}}\ln \frac{\left|
r-r_{-}\right| }{r_{-}}+c,
\end{equation}
where $\kappa _{+}$ and $\kappa _{-}$ have been defined by Eq. $\left( \ref
{p12ab}\right) $. To avoid singularities we can define Kruskal-Szekeres type
coordinates
\begin{equation}
V^{+}=\exp \kappa _{+}v\qquad U^{+}=-\exp -\kappa _{+}u.
\end{equation}
These coordinates do not cover $r\leq r_{-}$ because of the coordinate
singularity at $r=r_{-}$ (and $U^{+}V^{+}$ is complex for $r\leq r_{-}$),
but $r=r_{-}$ and a similar four regions are covered by the $\left(
U^{-},V^{-}\right) $ Kruskal-Szekeres-type coordinates to this case. Thus,
let us define
\begin{equation}
V^{-}=\exp \kappa _{-}v\qquad U^{-}=-\exp -\kappa _{-}u.
\end{equation}
For the $+$ sign we have
\begin{equation}
U^{+}V^{+}=-\exp \left( \kappa _{+}\left( v-u\right) \right) =-\exp \left(
2\kappa _{+}r^{\ast }\right) =-\exp \left( 2\kappa _{+}r\right) \left( \frac{%
r-r_{+}}{r_{+}}\right) \left( \frac{r-r_{-}}{r_{-}}\right) ^{\frac{\kappa
_{+}}{\kappa -}}
\end{equation}
and the respective line element is
\begin{equation}
ds_{+}^{2}=-\frac{r_{+}r_{-}}{\kappa _{+}^{2}}\frac{\exp \left( -2\kappa
_{+}r\right) }{r^{2}}\left( \frac{r_{-}}{r-r_{-}}\right) ^{\frac{\kappa _{+}%
}{\kappa -}-1}dU^{+}dV^{+}+r^{2}\left( U^{+},V^{+}\right) d\Omega ^{2},
\end{equation}
while for the $-$ sign we have
\begin{equation}
U^{-}V^{-}=-\exp \left( \kappa _{-}\left( v-u\right) \right) =-\exp \left(
2\kappa _{-}r^{\ast }\right) =-\exp \left( 2\kappa _{-}r\right) \left( \frac{%
r_{-}-r}{r_{-}}\right) \left( \frac{r_{+}-r}{r_{+}}\right) ^{\frac{\kappa
_{-}}{\kappa _{+}}}
\end{equation}
and the associated line element is
\begin{equation}
ds_{-}^{2}=-\frac{r_{+}r_{-}}{\kappa _{-}^{2}}\frac{\exp \left( -2\kappa
_{-}r\right) }{r^{2}}\left( \frac{r_{+}}{r_{+}-r}\right) ^{\frac{\kappa _{-}%
}{\kappa _{+}}-1}dU^{-}dV^{-}+r^{2}\left( U^{-},V^{-}\right) d\Omega ^{2}.
\end{equation}
The conformal Penrose diagram of the RN space is shown in Fig.\ref{f2}.
\section{The Hamiltonian contribution of the Electromagnetic field}
\label{app2}The form of the hamiltonian for the electromagnetic field can be
obtained with the same method used for the pure gravitational field. Let $%
u_{\alpha }$ a normalized time-like vector field s.t. $u_{\alpha }u^{\alpha
}=-1$. The form of ${\cal H}_{M}$ comes from the Einstein-Maxwell action and
it can be written as
\begin{equation}
{\cal H}_{M}=\sqrt{^{3}g}T_{\alpha \beta }u^{\alpha }u^{\beta },
\label{app21}
\end{equation}
where
\begin{equation}
T_{\alpha \beta }=\frac{1}{4\pi }\left[ F_{\alpha \gamma }F_{\beta }^{\gamma
}-\frac{1}{4}g_{\alpha \beta }F_{\gamma \delta }F^{\gamma \delta }\right]
\end{equation}
and $F_{\alpha \gamma }=\partial _{\alpha }A_{\gamma }-\partial _{\gamma
}A_{\alpha }$. $A_{\alpha }$ is the electromagnetic potential which, in the
case of a pure electric field assumes the form $A_{\alpha }=\left(
Q_{e}/r,0,0,0\right) $ while in the case of pure magnetic field, the form is
$A_{\alpha }=\left( 0,-Q_{m}\sin \theta ,0,0\right) $. $Q_{e}$ and $Q_{m}$
are the electric and magnetic charge respectively. Both of them contribute
in the same way to the gravitational potential of Eq. $\left( \ref{p12}%
\right) $. Since we are interested in electric type R.N. metrics, the
on-shell contribution of $T_{\alpha \beta }u^{\alpha }u^{\beta }$ is
\begin{equation}
T_{\alpha \beta }u^{\alpha }u^{\beta }=\frac{1}{8\pi }\left( F_{01}\right)
^{2}=\frac{1}{8\pi }\frac{Q_{e}^{2}}{r^{4}}.
\end{equation}
Nevertheless, this is not the complete contribution when gravitational
fluctuations come into play. Indeed up to second order in $h_{ij}$, one gets
\begin{equation}
\int_{\Sigma }d^{3}x{\cal H}_{M}=2\kappa \int_{\Sigma }d^{3}x\sqrt{^{3}\bar{g%
}}\left[ -\frac{1}{4}h_{ij}h^{ij}\left( T_{\alpha \beta }u^{\alpha }u^{\beta
}\right) +T_{\alpha \beta }^{\left( 2\right) }u^{\alpha }u^{\beta }\right] ,
\label{app22}
\end{equation}
where $T_{\alpha \beta }^{\left( 2\right) }u^{\alpha }u^{\beta }=\frac{1}{%
8\pi }\left( F_{01}\right) ^{2}h_{1}^{1}h_{1}^{1}$. Thus Eq. $\left( \ref
{app22}\right) $ becomes
\begin{equation}
\int_{\Sigma }d^{3}x{\cal H}_{M}=-\frac{2\kappa }{8\pi }\int_{\Sigma }d^{3}x%
\sqrt{^{3}\bar{g}}\left[ -\frac{3}{4}h_{1}^{1}h_{1}^{1}\left( F_{01}\right)
^{2}+\frac{1}{4}h_{2}^{2}h_{2}^{2}\left( F_{01}\right) ^{2}+\frac{1}{4}%
h_{3}^{3}h_{3}^{3}\left( F_{01}\right) ^{2}\right] .  \label{app23}
\end{equation}
On the other hand, when we consider the magnetic charge, the on-shell
contribution of $T_{\alpha \beta }u^{\alpha }u^{\beta }$ is
\begin{equation}
T_{\alpha \beta }u^{\alpha }u^{\beta }=\frac{1}{8\pi }\left( F_{23}\right)
^{2}=\frac{1}{8\pi }\frac{Q_{m}^{2}}{r^{4}}.
\end{equation}
However for the magnetic part, one gets
\begin{equation}
T_{\alpha \beta }u^{\alpha }u^{\beta }\varpropto
F_{ij}F^{ij}=F_{ij}F_{kl}g^{ik}g^{jl}
\end{equation}
then to second order in $h_{ij}$ one obtains
\begin{equation}
\int_{\Sigma }d^{3}x{\cal H}_{M}=-\frac{2\kappa }{16\pi }\int_{\Sigma }d^{3}x%
\sqrt{^{3}\bar{g}}\left[ \frac{1}{4}h_{1}^{1}h_{1}^{1}\left( F_{23}\right)
^{2}-\frac{7}{4}h_{2}^{2}h_{2}^{2}\left( F_{23}\right) ^{2}-\frac{7}{4}%
h_{3}^{3}h_{3}^{3}\left( F_{23}\right) ^{2}\right] .
\end{equation}
\section{Action and Hamiltonian}
\label{app3}Here we follow Ref.\cite{FroMar} to extract the form of the
Hamiltonian in presence of a bifurcation surface with boundaries. We will
consider for simplicity the case of a single horizon, like the Schwarzschild
one, the generalization for the Reissner-Nordstr\"{o}m case is
straightforward. We consider two spacelike Cauchy surfaces $\Sigma ^{\prime
} $ and $\Sigma ^{\prime \prime }$. The region lying between $\Sigma
^{\prime } $ and $\Sigma ^{\prime \prime }$ consists of two wedges $M_{+}$
and $M_{-}$ intersecting at a two-dimensional surface $S_{0}$. The symbol $%
\Sigma _{t\,(\pm )}$ denotes the part of $\Sigma _{t}$ located in $M_{\pm }$%
. We also denote those parts of $\Sigma ^{\prime }$ and $\Sigma ^{\prime
\prime }$ which are the spacelike boundaries of the wedges $M_{\pm }$ as $%
\Sigma _{\pm }^{\prime }$ and $\Sigma _{\pm }^{\prime \prime }$. The lapse
function $N$ is positive (negative) at $M_{+}$ ($M_{-}$) and equals zero at
the bifurcation surface. The vector $u^{\mu }=-N\partial _{\mu }t$ is future
oriented in $M_{+}$ and past oriented in $M_{-}$. The metric is described in
Eq.$\left( \ref{p14}\right) $ and the covariant form of the gravitational
action for this foliation with fixed three-geometry at the boundaries of $M$
is
\[
S={\frac{1}{2\kappa }}\int_{M_{+}}d^{4}x\,\sqrt{-g}\,\Re +{\frac{1}{\kappa }}%
\int_{t^{\prime }({+})}^{t^{\prime \prime }}d^{3}x\,\sqrt{g^{\left( 3\right)
}}\,K-{\frac{1}{\kappa }}\int_{{B_{+}}}d^{3}x\,\sqrt{-\gamma }\,\Theta
-S_{0\left( +\right) }
\]
\begin{equation}
-{\frac{1}{2\kappa }}\int_{M_{-}}d^{4}x\,\sqrt{-g}\,\Re +{\frac{1}{\kappa }}%
\int_{t^{\prime }({-})}^{t^{\prime \prime }}d^{3}x\,\sqrt{g^{\left( 3\right)
}}\,K-{\frac{1}{\kappa }}\int_{{B}_{-}}d^{3}x\,\sqrt{-\gamma }\,\Theta \
-S_{0\left( -\right) }.  \label{app31}
\end{equation}
$\Re $ denotes the four-dimensional scalar curvature, $\kappa \equiv 8\pi G$%
, and
\[
\sqrt{-g}=|N|\sqrt{h}
\]
\begin{equation}
\sqrt{-{\gamma }}=|N|\sqrt{\sigma },
\end{equation}
where $\sigma $ is the determinant of the two-dimensional metric $\sigma
_{ab}$. $K$ is the extrinsic curvature of $\Sigma _{t}$ as a surface
embedded in $M$, while $k$ is the extrinsic curvature of the boundaries $S$
embedded on $\Sigma $. Then the trace of the extrinsic curvature of the
boundaries $B$ as surfaces embedded in $M$ is
\begin{equation}
\Theta =k-n_{\beta }a^{\beta }\ ,  \label{app32a}
\end{equation}
where $a^{\mu }=u^{\alpha }\nabla _{\alpha }u^{\mu }$ is the acceleration of
the timelike normal $u^{\mu }$. $B$ is a three-dimensional timelike boundary
such that $B=B_{+}\cup B_{-}$. The spacelike normal $n^{i}$ to the
three-dimensional boundaries $B$ is assumed to be outward pointing at $B_{+}$%
, inward pointing at $B_{-}$. It is assumed that the integrations
are taken over the coordinates $x^{\mu }$ which have the same
orientation as the canonical coordinates $(t,x,\theta ,\phi )$ of
the foliation. The negative sign for the integration over $M_{-}$
reflects the fact that the canonical coordinates are left oriented
in this region. Besides the volume
term, the action $S$ contains also boundary terms. The notation $%
\int_{t^{\prime }({\pm })}^{t^{\prime \prime }}$ represents an integral over
the three-boundary $\Sigma _{\pm }$ at $t^{\prime \prime }$ minus an
integral over the three-boundary $\Sigma _{\pm }$ at $t^{\prime }$. Under a $%
3+1$ spacetime split, the four-dimensional scalar curvature is
\begin{equation}
\Re =R^{\left( 3\right) }+K_{\mu \nu }K^{\mu \nu }-(K)^{2}-2\nabla _{\mu
}(Ku^{\mu }+a^{\mu })\ ,  \label{app33}
\end{equation}
where $R$ denotes the scalar curvature of the three-dimensional spacelike
hypersurface $\Sigma $. By the use of Gauss' theorem, the conditions
\begin{equation}
u\cdot n|_{B}=0,\,u\cdot a=0,\,u\cdot u=-1,\,n\cdot n=1\ ,
\end{equation}
and Eqns. (\ref{app31})-(\ref{app32a}) and (\ref{app33}), one can
rewrite the total action in the form
\[
S={\frac{1}{2\kappa }}\int_{M_{+}}d^{4}x\sqrt{-g}\,\left[ R+K_{\mu \nu
}K^{\mu \nu }-(K)^{2}\right] -{\frac{1}{\kappa }}\int_{{B_{+}}}d^{3}x\sqrt{%
-\gamma }\,k
\]
\begin{equation}
-{\frac{1}{2\kappa }}\int_{M_{-}}d^{4}x\sqrt{-g}\,\left[ R+K_{\mu \nu
}K^{\mu \nu }-(K)^{2}\right] -{\frac{1}{\kappa }}\int_{{B_{-}}}d^{3}x\sqrt{%
-\gamma }\,k\ .  \label{app34}
\end{equation}
The action for stationary solutions finally becomes
\begin{equation}
S=-\int Hdt\ ,
\end{equation}
with the gravitational Hamiltonian $H$ given by
\begin{equation}
H=\int_{\Sigma }d^{3}x\left( N{\cal H}+N^{i}{{\cal H}_{i}}\right) +\frac{1}{%
\kappa }\int_{{S_{+}}}d^{2}x\sqrt{\sigma }Nk-\frac{1}{\kappa }\int_{{S_{-}}%
}d^{2}x\sqrt{\sigma }Nk.  \label{app35}
\end{equation}

\end{document}